\documentclass{aa}  

\usepackage{graphicx}
\usepackage{epsfig}
\usepackage{txfonts}
\usepackage{natbib}
\begin{document} 
\newcommand{\kms}{km s$^{-1}$}

   \title{Spectral decomposition of the stellar kinematics in the polar disk
galaxy NGC 4650A}

   \author{L. Coccato \inst{1}
          \and
          E. Iodice \inst{2}
          \and
          M. Arnaboldi \inst{1}
          }

   \institute{ESO,  Karl-Schwarzschild-Strasse 2, D-85748 Garching, Germany\\
         \and
             INAF-Astronomical Observatory of Capodimonte, via
     Moiariello 16, I-80131 Naples, Italy\\
                  }

   \date{Received .....; accepted .....}

 
  \abstract
  {The prototype of Polar Ring Galaxies NGC 4650A contains two main
    structural components, a central spheroid, which is the host galaxy,
    and an extended polar disk. Both photometric and kinematic
    studies revealed that these two components co-exist on two
    different planes within the central regions of the galaxy.}
   {The aim of this work is to study the spectroscopic and kinematic
     properties of the host galaxy and the polar disk in the central
     regions of NGC 4650A by disentangling their contributions to the 
     observed galaxy spectrum.}
   {We applied the spectral decomposition technique introduced in
     previous works to long-slit spectroscopic observations in the
     Ca{\small II} triplet region.  We focused the analysis along the
     $P.A. = 152^{\circ}$ that corresponds to the photometric minor
     axis of the host galaxy, where the superimposition of the two
     components is more relevant and the spectral decomposition is
     necessary. We aim to disentangle the stars that move in the
     equatorial plane of the host galaxy from those that move in the
     meridan plane, which is along the polar disk.}
   {We successfully disentangled the spectra of the two structural
     components of NGC 4650A and measured  their line-of-sight velocity and
     velocity dispersion profiles, and the stellar content along
     $P.A. = 152^{\circ}$. The host galaxy shows significant rotation
     along its photometric minor axis, indicating that the
     gravitational potential is not axisymmetric. The polar disk shows
     a kinematic decoupling: the inner regions counter-rotating
     with respect the outer-regions and the host spheroid. This
     suggests a complex formation history for the polar disk,
     characterised by mass accretion with decoupled angular momenta.}
{}

   \keywords{Galaxies: kinematics and dynamics; Galaxies: individual: NGC~4650A}

   \maketitle
%

\section{Introduction}
\label{sec:intro}

\begin{figure}
  \psfig{figure=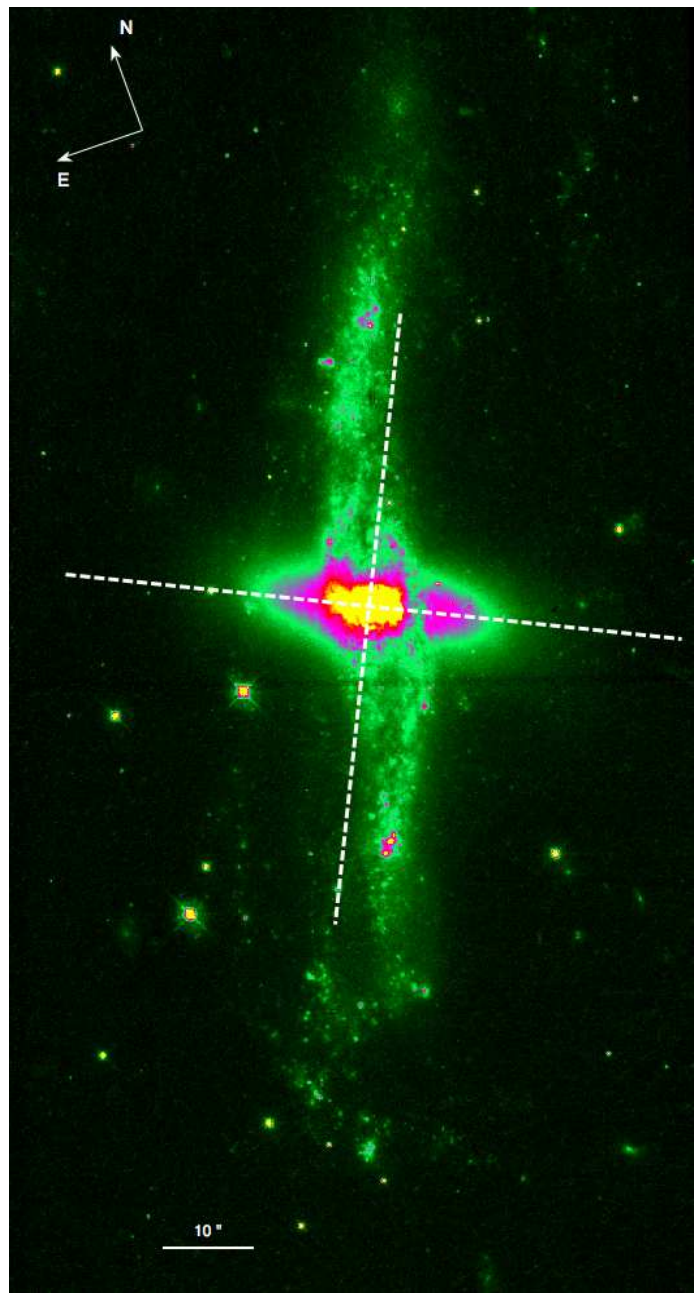,width=9cm,clip=}
  \caption{NGC~4650A in the optical F450W HST band
    \citep{Iodice+02}. Image size is $1.7 \times 2.4$~arcmin. The
    white dashed lines indicate the directions of the FORS2 long slit
    spectra acquired by \citet{Iodice+06}, along the major ($P.A. =
    62^{\circ}$) and minor ($P.A. = 152^{\circ}$) axis of the host
    galaxy. East and north directions are indicated on the image.}
\label{4650a}
\end{figure}

The galaxy NGC~4650A is a polar disk galaxy (see Fig.~\ref{4650a}). It
is one of the best-investigated objects among the class of Polar Ring
Galaxies (PRGs). The PRGs are multi-spin systems composed of a central
spheroid (which is the host galaxy), and a polar structure, which
orbits nearly perpendicularly to the equatorial plane of the host
galaxy. The latest studies on PRGs have revealed that this
morphological type of galaxies includes both polar rings and polar
disks, which have a different structure (i.e., light distribution,
colors, age, and kinematics) and, probably, a different formation
history \citep[see][for a review]{Iodice2014}.

{\it Why is it interesting to study PRGs?} The multi-spin morphology
of PRGs cannot be explained by the collapse of a single proto-galactic
cloud, but some kind of interaction (galaxy-galaxy or
galaxy-environment) needs to be invoked in the formation history of
these systems. The gravitational interactions are the `carrying
pillar' of the cold dark matter model for galaxy formation, and in this
framework, both merger and gas accretion play a major role in building
the structure of spheroid and disk \citep[see][as
review]{Conselice2014}. Thus, given their unique geometry, PRGs are
among the best objects in the universe to study the physics of such
processes.  Moreover, the existence of two orthogonal components of
the angular momentum makes the PRGs the ideal laboratory to derive the
three-dimensional shape of the gravitational potential \citep[see][as
reviews]{Combes2014, Arnaboldi2014}.

{\it How could a PRG form?} To date, three main formation scenarios
have been proposed for PRGs: 1) a major dissipative polar merger of
two disk galaxies with unequal mass \citep{Bekki1998, BC2003}; 2) the
tidal accretion of external material (gas and/or stars), captured by
an early-type galaxy on a parabolic encounter \citep{RS1997, BC2003,
  Hancock2009}; 3) the cold accretion of pristine gas along a filament
\citep{Maccio2006, Brook2008}. As for any galaxy formation mechanism,
the proposed scenarios need to account for the morphology, gas
content, star and gas kinematics, and stability. Therefore, the key
physical parameters that allow discriminating among the three
formation scenarios are 1) the baryonic mass (stars plus gas) ratio
between the host galaxy and polar structure; 2) the kinematics and
orbital distributions of the stars in the host galaxy; 3) the
metallicity and star formation rate in the polar structure.

In the last ten years, extensive observational campaigns were carried
out to derive the above quantities for NGC 4650A and, thus, to outline
the structure and formation history of this fascinating and intriguing
galaxy.  Optical (HST $B$, $V$, and $I$ bands) and near-infrared ($J$,
$H$, and $Ks$ bands) photometry, emission and absorption line long-slit
spectroscopy and HI radio emission showed that the polar structure is
a disk, rather than a ring, because its stars and dust can be traced
inward within the host galaxy, down to $\sim$~1.2~kpc from the galaxy
nucleus \citep{Arnaboldi1997, Iodice+02, Gallagher2002, Swaters2003}.
The polar disk contains a large amount of HI ($\sim
10^{11}$~M$_{\odot}$), which is four times more extended than the
optical stellar disk (out to 40~kpc from the galaxy center). Thus the
baryonic mass in the polar structure is equal or even larger than that
in the host galaxy, which is a gas-free object. The polar disk has a
sub-solar metallicity $Z = 0.2$~Z$_{\odot}$, and there is no
metallicity gradient along this component \citep{Spavone+2010}.  The
host galaxy in NGC~4650A has morphology, light distribution, and colors
that resemble those of an S0 galaxy.  Optical long-slit spectra along
the major and minor axis of the host galaxy \citep{Iodice+06} have
confirmed that this component is rotationally-supported with a
maximum rotation velocity along the major axis of $v_{rot} \simeq 80 -
100$~km~s$^{-1}$. Some rotation is measured also along the minor axis
outside of 1 kpc.  The velocity dispersion remains almost constant
($\sigma \sim 65$~km/s) at all radii and along both axes. These
measurements put the host galaxy in NGC~4650A far from the
Faber-Jackson relation for early-type galaxies
\citep[see][]{Iodice+06, Iodice2014}, and show lower central velocity
dispersion with respect to spheroids of comparable luminosity.  From
dynamical studies on NGC~4650A, the best models predict a flattened
E6-E7 dark halo with its major axis aligned along the plane of the
polar disk itself \citep{CA1996, Nap2014}.

 In summary, the collected data, and the overall structure and
  morphology indicate that NGC~4650A is consistent with being a polar
  disk galaxy formed through the accretion of external cold gas from
  cosmic web filaments \citep[see][]{Spavone+2010}.

 Nevertheless, the kinematic measurements and the dynamical models
  of NGC 4650A contain two major sources of uncertainty. The first is
  the limitation of 1D long-slit spectra that cannot trace the full
two-dimensional warped geometry of the polar disk. The second is
the contamination of the kinematics of the host galaxy by the stars in
the polar disk, in the regions where the two components coexist. 
Indeed, the stellar kinematics of the host galaxy along its
photometric minor axis ($P.A. = 152^{\circ}$) are not what is expected
for axisymmetric systems. In the inner 1 kpc, the velocity dispersion
profile is slightly asymmetric with respect to the galaxy
center. Moreover, outside 1 kpc, the velocity dispersion rise up to
120 \kms\, and the rotation velocity increases up to $\sim 70$ \kms.
Previous analysis was not able to give a unique interpretation for
these effects, and suggested three viable explanations
\citep{Iodice+06} (i) a non axisymmetric structure of the host galaxy
in NGC~4650A; (ii) a contamination by the stellar motion in the polar
disk, or (iii) a contamination by stronger Paschen lines in a mixed stellar
population.

In this paper, we investigate the scenario where the polar disk
  contaminates the kinematics of the host galaxy. Our aim is to
separate the stellar kinematics of the two components and, therefore,
to remove their mutual contamination on the observed velocity
distribution along the line of sight -- by applying a spectral
decomposition technique to the available FORS2 long-slit spectra
\citep{Iodice+06}.

In this paper, we adopt a distance to NGC 4650A of 43.4 Mpc and
$H_0=73$~km~s$^{-1}$~Mpc$^{-1}$, which implies a scale of 0.21 kpc
arcsec$^{-1}$ (from the NASA/IPAC Extragalactic Database).

This paper is structured as follows: in Sec.~\ref{sec:analysis}, we
describe the analysis performed on the long slit spectra; results are
discussed in Sec.~\ref{sec:spec_dec_results} ,and concluding remarks
are drawn in Sec.~\ref{concl}.

\section{Analysis}
\label{sec:analysis}

As stated in the introduction, the aim of this work is to disentangle
the relative contribution of the polar disk towards the inner regions
of the host galaxy in NGC~4650A, where the two structural components
overlap along the line of sight. To this aim, we apply the spectral
decomposition technique described in \citet{Coccato+11, Coccato+13},
which is an implementation of the pPXF fitting code
\citep{Cappellari+04}, to the FORS2 spectra \citep{Iodice+06}.

The spectra were obtained with FORS2 at the UT4 of the ESO Very Large
Telescope (Chile) in service mode. The FORS2 was equipped with the grism
GRIS 1028z+29 and a long-slit 1\farcs6 wide. The spectral resolution
was $\sigma \sim 70$ \kms\ measured at 8600 \AA. We refer to
\citet{Iodice+06} for more information on the instrumental set-up,
data acquisition, and reduction.

\subsection{Spectral decomposition}

The spectral decomposition code builds two optimal templates as a
linear combination of stellar spectra from two input libraries. These
two optimal templates are convolved by two Gaussian functions with
independent velocities and velocity dispersions. The two convolved
optimal templates are then normalised to their median values within a
desired wavelength range $W$, and multiplied by two factors $F_1$ and
$F_2=T-F_1$, where $T$ is the median flux of the observed galaxy
spectrum in the wavelength range $W$.  These steps are embedded in a
$\chi^2$ minimisation loop till convergence. Multiplicative
polynomials that account for the shape of the continuum are included
in the process.
The spectral decomposition code returns the velocities ($V_1, V_2$),
velocity dispersions ($\sigma_1, \sigma_2$), the flux contribution
$F_1$ of the first component, and the best fitting templates of the
two components.

In our fitting process, we define the wavelength $W$ to be $8500 \AA <
\lambda < 8800 \AA$ and normalise the input galaxy spectra, so that
$T=1$. The parameters $F_1$ and $F_2$ are then the fractional
contributions to the median galaxy flux of the hosting galaxy and the
polar disk, respectively.

The accuracy of the spectral decomposition results and the ability to
break the degeneracies between the various parameters (i.e., flux ratio
of the two components, the choice of the stellar templates, and the
kinematics) depend on how different the stellar components are, on the
instrumental set-up, and on the observed signal-to-noise ratio.  In
our case, the limiting parameter for the analysis is the wavelength
range, which contains only few spectral features (Ca{\small II}
triplet).

Ad-hoc assumptions can be made to decrease the number of free
parameters in the fit and to reduce degeneracies between the various
parameters in the spectral decomposition process. \citet{Katkov+13}
and \citet{Fabricius+14} adopted a non parametric kinematic fit, which
was used to constrain the kinematics of the two components and fit
their stellar populations on a later stage. In our work, we adopt a
different approach, following the prescriptions listed below.

\begin{enumerate}

  \item Constrain the stellar template of the host galaxy using
    information on regions where the disk contribution is negligible
    (Section \ref{sec:galaxy_template}).

  \item Constrain the flux ratio between the host galaxy and the polar
    disk  using the light profile obtained by collapsing the
      long-slit spectra along the dispersion direction (Section
    \ref{sec:flux_ratio}).

\end{enumerate}

In Section \ref{sec:spec_dec_results}, we present the results of the
spectral decomposition on NGC~4650A.

\subsection{Host galaxy template}
\label{sec:galaxy_template}

Spectra obtained along the major axis of the host galaxy
($P.A.=62^{\circ}$) on the NE side of NGC~4650A, and outside
the inner $\sim 3"<R<10"$ do not contain any relevant disk contribution (see
Fig.~\ref{4650a}). Therefore, they can be used to obtain the stellar
template that describes the mean properties of the stellar population
of the host galaxy.

To extract best stellar template of the host galaxy, we fit the
radially binned galaxy spectra with a selection of stars from the
Ca{\small II} triplet library of \citet{Cenarro+01}, which have a
spectral resolution of 1.5\AA (FWHM). The stellar library contains 88
giant stars that sample the $\log g$, $T_{eff}$, and [Fe/H] parameter
space with spectral types ranging from O to M.  We broadened the
stars in the library to match the instrumental Line Spread Function
(LSF) of our FORS2 observations. The LSF has been parametrized with a
Gaussian Function and higher order Hermite Polynomials ($\sigma=2.0
\AA$, $h3=0.0$, $h4=-0.16$), as measured from the comparison arc
spectra in the regions of the detector relevant for our analysis. In
these regions, we did not observe significant variations of the LSF.

The fit was performed by setting the contribution of the disk
component in the spectral decomposition code to zero. The code
determined a best fitting template as a linear combination of the stars
in the library for each radial bin. These best fitting templates are
very similar (variations have a standard deviation of 0.7\%),
suggesting a negligible radial dependence of the stellar populations
in the host galaxy.  We therefore defined the host galaxy template as
the mean spectrum of the best fitting templates at different
radii. Having constrained the host galaxy stellar template will reduce
considerably the degeneracies in the spectral decomposition code.

\begin{figure}
  \psfig{figure=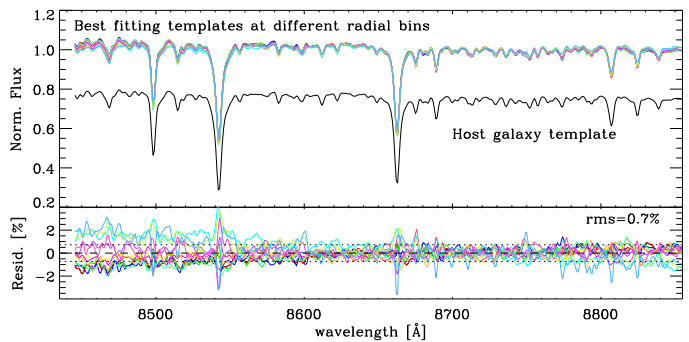,width=9cm,clip=}
  \caption{Upper panel: spectra of the best fitting templates
    determined at different radial bins (along the major axis of the
    host galaxy at $PA=62^{\circ}$;  3$"<R<10"$, NE side of the
    galaxy. See the slit orientation shown in Fig.~\ref{4650a}). The
    average of these spectra is shown in black (shifted by $-0.2$ for
    clarity purposes); it is be used as template spectrum for the
    host galaxy in the spectroscopic decomposition (see text for
    details). Lower panel: deviations of the templates from the mean
    spectrum.}
\label{fig:galaxy_templates}
\end{figure}

In Fig.~\ref{fig:galaxy_templates} we compare the best fitting
templates for the various radial bins and their mean spectrum, which
we use to define the host galaxy template.  The best-fitting templates
are composed mainly of GIII ($\sim$ 50\%, in light) and KIII ($\sim$
35\%) giant stars and a small contribution from young O, B, and A
stars ($\sim$ 10\%) and MIII giants ($< 5$\%).  The lack of strong
Balmer absorption line features in the studied wavelength range, such
as H$\beta$, prevented a more accurate determination of the age of the
stellar population.

\subsection{Flux ratio between spheroid and disk}
\label{sec:flux_ratio}

In this section, we exploit the different surface brightness radial
profiles of the host galaxy and the polar disk along the slit at
$P.A.=152^{\circ}$ to reduce the degeneracies between their fluxes,
kinematics, and stellar populations when applying the spectral
decomposition. The use of photometric constraints to spectral
decomposition techniques has been successfully applied also in
previous works (i.e., \citealt{Johnston+12}).
  
According to \citet{Iodice+02}, we model the light profile of the
long-slit at $P.A.=152^{\circ}$ with the sum of two components, a
Sersic law for the host galaxy, $G(R)$, and an exponential law,
$D(R)$, to model the light distribution in the polar disk.  The host
galaxy is parametrized by

\begin{equation}
G(R) = \mu_e + k(n) \left[ \left( \frac{R}{r_e}\right)^{1/n} -1\right],
\end{equation}
where $R$ is the galactocentric distance along $P.A.=152^{\circ}$,
$r_e$ and $\mu_e$ are the effective radius and effective surface
brightness respectively, and $k(n)=2.17 n - 0.355$. The polar disk is
parametrized by

\begin{equation}
D(R)=  \mu_0 + 1.086 \left( \frac{R}{r_h}\right),
\end{equation}
where $\mu_0$ and $r_h$ are the central surface brightness and
scalelength of the polar disk.

We applied a photometric offset of $\Delta \mu = 33.8$~mag to the
instrumental magnitudes measured from the long-slit profile, so that
the surface brightness profile matches the one derived from the HST
image of NGC~4650A in the F814W filter (see Fig. 6 in
\citealt{Iodice+02}).

The best fit of the scaled light profile gives the following values
for the above structural parameters: $\mu_e= 19.30 \pm
0.08$~mag/arcsec$^2$, $r_e = 2.00 \pm 0.05$~arcsec, $n = 1.2 \pm 0.1$,
$\mu_0 = 20.20 \pm 0.06$~mag/arcsec$^2$, and $r_h = 5.8 \pm 0.1$~arcsec.

\begin{figure}
  \psfig{figure=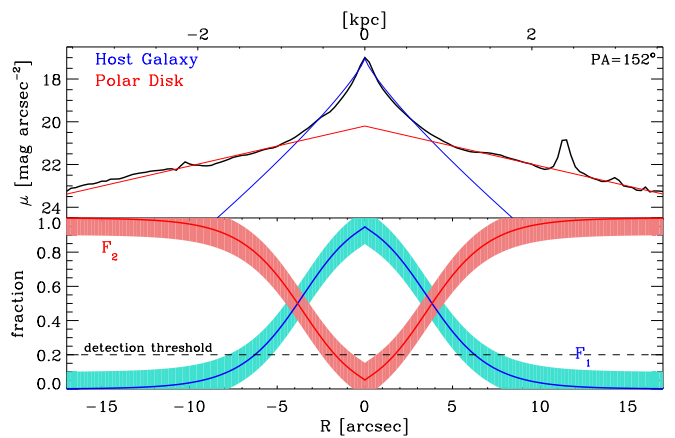,width=9cm,clip=,bb=40 22 580 367}
  \caption{Light contribution of the host galaxy and the polar
    disk. Upper panel: the long-slit luminosity profile (instrumental
    surface brightness) obtained by collapsing the long-slit spectrum
    ($PA=152^{\circ}$) along the wavelength direction in the $8500<
    \lambda< 8800$ region. The profile is decomposed into the
    contribution of a Sersic profile (associated to the host galaxy,
    in blue) and an exponential disk (associated to the polar disk, {\bf in
    red}). Bottom panel: fractional contribution of the host galaxy
    ($F_1$, in blue) and the polar disk ($F_2$, in red) with respect
    to the median galaxy flux (normalized to 1). For a given radial
    bin $R$, the spectral decomposition code computes the best fit
    value $F_1^{\rm bestfit}$ in the interval $F_1(R)-0.1 < F_1^{\rm
      best fit} < F_1(R)+0.1$ (which is represented by the shaded
      regions in the lower panel), and it assumes $F_{1,2}=0.2$ as
    detection limit for a component (represented by the dashed
    horizontal line). See Section~\ref{sec:spec_dec_results} for
    details.}
\label{fig:flux_ratio}
\end{figure}

In the fit, we considered the same wavelength range used in the
spectral decomposition ($8500 \AA < \lambda < 8800 \AA$).
Figure ~\ref{fig:flux_ratio} shows the results of the photometric
decomposition.

At each radial bin $R$, we use the photometric decomposition to set a
starting guess value of $F_1$ as

\[
F_1(R) = \frac{G(R)}{G(R) + D(R)}.
\]
In our convention, component 1 is the host galaxy, whereas component 2
is the polar disk.  We have $F_2(R) = 1-F_1(R)$ by construction at
each $R$.

\section{Results of the spectroscopic decomposition}
\label{sec:spec_dec_results}

Following \citet{Iodice+06}, we radially binned the spectrum along
$PA=152^{\circ}$ to reach a minimum $S/N$ ratio of 50 per pixel. At
each radial bin $R$, we applied the spectroscopic decomposition, fixing
the stellar template of the host galaxy to the optimal template
determined in Section \ref{sec:galaxy_template} and using a selection
of O, A, B, G, K, and M stars from the \citet{Cenarro+01} library for
the polar disk. At each radial bin, the flux contribution of the host
galaxy is given by $F_1(R)$ (see Section~\ref{sec:flux_ratio}); the
best fitting value $F_1^{\rm best fit}$ is searched in the $F_1(R)-0.1
< F_1^{\rm best fit} < F_1(R)+0.1$ range. The amplitude of this range
is consistent with the errors on the photometric decomposition as
described in Section~\ref{sec:flux_ratio}. A detection threshold of
$F_{1,2}=0.2$ is used for each component (see
Section~\ref{sec:errors}).

\begin{figure*}
\centering
  \psfig{figure=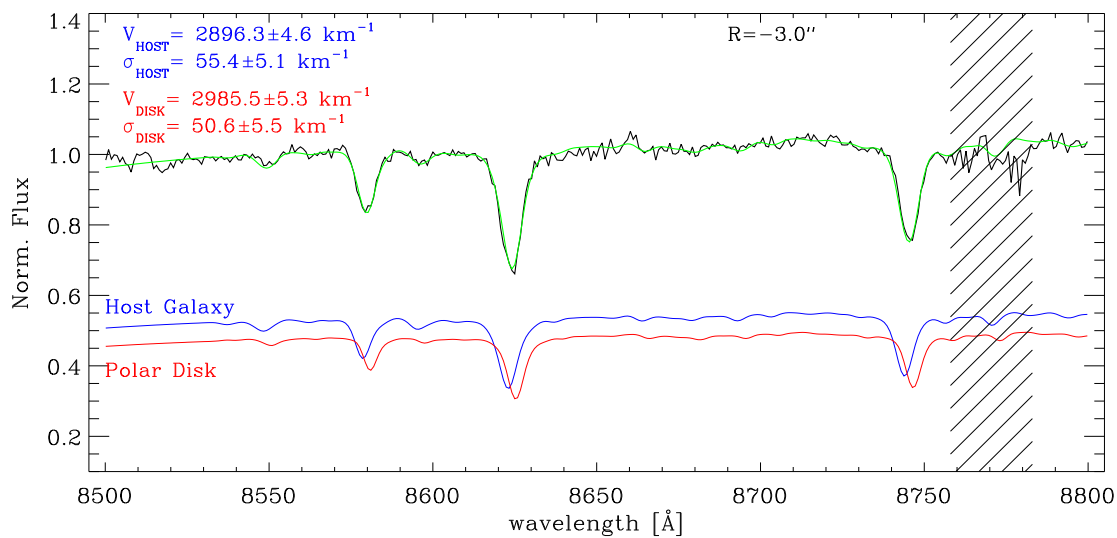,width=15cm,clip=}
\caption{Example of results of the spectroscopic decomposition at a
  given location ($R=-3''$, South east, $P.A.=152^{\circ}$). The observed
  spectrum of NGC 4650A (black) is decomposed into the contribution of
  the polar disk (red) and the host galaxy (blue). Regions excluded
  from the fit are shaded by diagonal black lines. The best fit model is shown in
  green.}
\label{fig:result_fit}
\end{figure*}

As an example, we show the result of the spectroscopic decomposition
  in one radial bin in Figure \ref{fig:result_fit}.

\subsection{Kinematics}

\begin{figure*}
\centering
  \psfig{figure=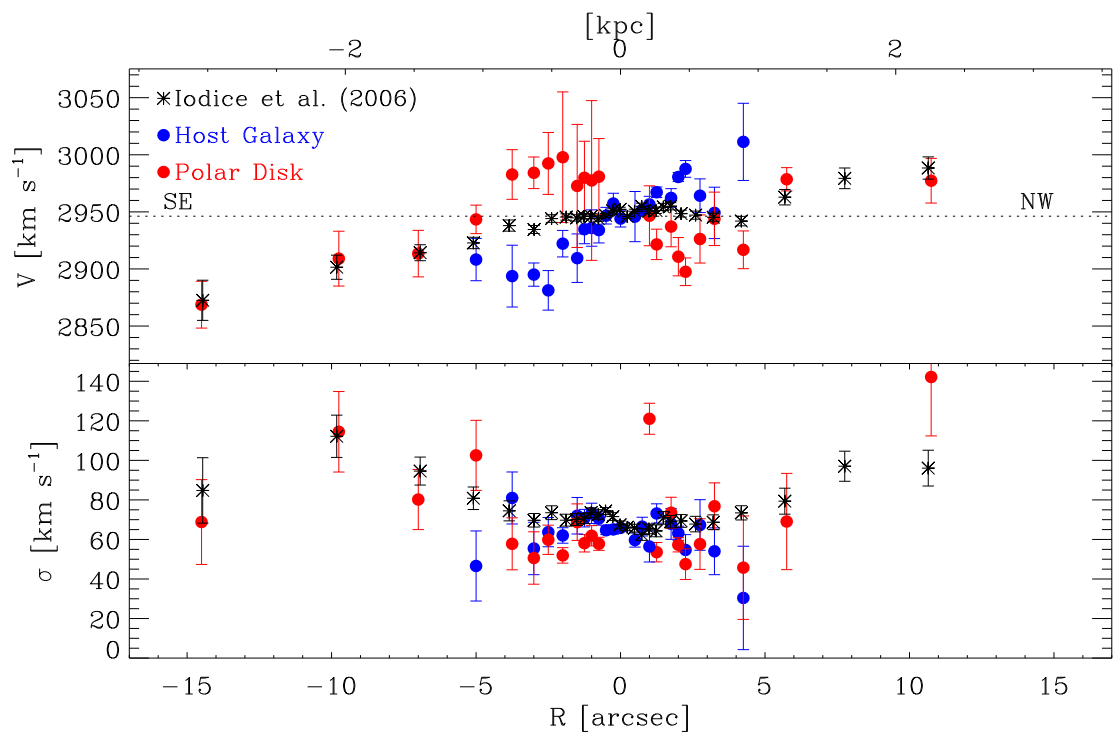,width=15cm,clip=}
\caption{Stellar rotation curves (top) and velocity dispersion
  profiles (bottom) of the two kinematics components in NGC 4650A
  measured along $P.A.=152^{\circ}$. Measurements referring to the
  host galaxy and the polar disk are shown by the blue and red
  circles, respectively. Kinematics of components whose fractional
  flux contribution $F_{1,2}$ is below the detection limit 0.2 are not
  shown. Results from a single component fit \citep{Iodice+06} are
  shown for comparison (black asterisks).}
\label{fig:result_rot_curve}
\end{figure*}

Figure~\ref{fig:result_rot_curve} shows the rotation curves and the
velocity dispersion radial profiles of the host galaxy and the polar
disk, along the host galaxy  photometric minor axis. The
contribution of the host galaxy to the total galaxy spectrum is
negligible outside 5~arcsec ($F_1^{\rm best fit}(|R|>5'') < 0.2$),
whereas the contribution of the polar disk is negligible inside $1''$
($F_2^{\rm best fit}(|R|<1'') < 0.2$). In the intermediate regions
$1''<|R|<5''$, the two components co-exist and the spectroscopic
decomposition is needed to separate and measure their
  properties independently.

The rotation curve of the polar disk is ``S''-shaped with the central
$\sim 3$~arcsec ($\sim 0.6$~kpc) counter-rotating with respect the
outer regions. The maximum amplitude of the rotation is $\sim 70$
\kms, which is measured at $15'' (\sim 3)$~kpc. The maximum amplitude of the
counter-rotation is $\sim 50$ \kms, which is measured at $2" (\sim 0.4)$~kpc.

Rotation of $\sim 70$ \kms\ is observed also for the host
galaxy. Since the observed position angle $P.A.=152^{\circ}$
corresponds to the host galaxy photometric minor axis, the observed
rotation indicates that the gravitational potential is not
axisymmetric.

The line-of-sight velocity dispersion $\sigma_{\rm LOS}$ within
$5''(\sim 1)$~kpc is $\sigma_{\rm LOS} \sim 60$\kms\ for both
components. In the outer regions, the velocity dispersion of the polar
disk increases up to $\sigma_{\rm LOS} \approx 120$ \kms.  The
contamination from the host galaxy in these regions is below the
detection limit, which is $<20$\%. This implies that the polar disk is
dynamically hot, if compared to typical disks of spiral galaxies.

Unfortunately, the $S/N$ ratio of our spectra is not enough to
reliably measure higher order moments of the velocity distributions
($h3$ and $h4$). Indeed, when including higher moments in the LOSVD
parametrization, the fit procedure became unstable. The general shapes
of the rotation curves shown in Figure \ref{fig:result_rot_curve} did
not change dramatically, but the higher moments were jumping from
positive to negative values with a strong dependence from the
starting guesses; this made it impossible to resolve the degeneracy
between the kinematic parameters of the two components. The non
inclusion of $h3$ and $h4$ in the fit could have introduced a bias in
our results towards higher rotational velocities, or different
$\sigma$ (e.g. higher/lower estimate of $\sigma$ for negative/positive
$h4$). Nevertheless, we believe that this bias is small: our
measurements agree well with those of \citet[in which a single
component fit was performed]{Iodice+06} in the regions where one
component dominates (with the exception of the outermost measurement
of $\sigma$ at 12'').

\subsection{Detection threshold and errors on kinematics}
\label{sec:errors}

Errors on kinematics are computed by means of Monte Carlo
simulations. For each radial bin $R_{\rm bin}$, we constructed a set
of 100 simulated spectra by combining the best fit models of the host
galaxy and the polar disk. The light contributions of the two components
to the simulated spectrum are given by $F_1(R_{\rm bin})$ (host
galaxy) and $1-F_1(R_{\rm bin})$ (polar disk), as defined in
Section~\ref{sec:flux_ratio}. We added a Gaussian noise to match the
$S/N$ measured at $R_{\rm bin}$, and a systematic component that
accounts for residuals from sky subtraction.

We then performed the spectroscopic decomposition on the 100 simulated
spectra and took the standard deviations of the 100 measurements of
each kinematic quantity (i.e., $V_1, V_2, \sigma_1,$ and $\sigma_2$) as
error. Errors are shown in Figure~\ref{fig:result_rot_curve}.

In the simulations, the velocity error of a component that contributes
less than 20\% to the total galaxy light is $\Delta V \geq 100$
\kms. We therefore set $F_{1,2} = 0.2$ as a detection limit for each
component (represented by the dashed horizontal line in
Figure~\ref{fig:flux_ratio}). This limit is consistent with the one
found in \citet{Coccato+13}.

\subsection{Stellar content of the polar disk}

If we consider the entire radial range, the best-fitting templates of
the disk are composed mainly of GIII ($\sim$ 45\%, in light) and KIII
($\sim$ 35\%) giant stars with contamination from young O, B, and A
stars ($\sim15$\%), and MIII giant stars ($<5$\%). The contribution of
young stars has a radial gradient: it is lower in the inner 1 kpc (up
to $\sim$ 10\% of the galaxy flux for $R\leq6''$) and higher in the
outer regions (up to $\sim$ 30\% for $R>6''$).

Our estimates for the stellar type content in the polar disk are
slightly lower than those reported by \citet{Iodice+06}, which
estimated a contamination of young stars up to 50\% for $R>6''$ (along
the same position angle). The difference arises from the different
analysis and the number of stars used in the stellar library.  We
evaluate the fraction (in light) of young stars by using the entire
available wavelength range ($8500\AA < \lambda < 8800\AA$), whereas
they considered only the Pa(17-14) versus Ca{\small II} equivalent
width ratio. Moreover, they used one A and one K template star,
whereas we used the \citet{Cenarro+01} library.

\section{Conclusions}\label{concl}

We studied the kinematics and stellar content of the two main
components in NGC~4650A, which are the central spheroid (host galaxy)
and the polar disk (see Fig.~\ref{4650a}). These two components
coexist in the central region of the galaxy; therefore, a careful
spectral decomposition is needed to isolate their respective
contributions to the observed galaxy spectrum. To this aim, we applied
the spectroscopic decomposition developed in \citet{Coccato+11} to the
long-slit spectroscopic observations of \citet{Iodice+06} obtained
with FORS2 at the Very Large Telescope.

We focused the analysis along the photometric minor axis of the host
galaxy ($P.A. = 152^{\circ}$), where the superposition of the two
components is more relevant.  The main results are the following:

\begin{itemize}

\item {\it Rotation.} Both components, the host galaxy and
  the polar disk, rotate along $P.A. = 152^{\circ}$. The rotation
  curve of the polar disk is ``S''-shaped with the central $\sim
  0.6$~kpc counter-rotating with respect to the outer regions and the
  host galaxy.

\item {\it Velocity dispersion.} The velocity dispersion within $\sim
  1$~kpc is $\sim 60$ \kms\ for both components. In the outer regions,
  the velocity dispersion of the polar disk increases up to $\approx
  120$ \kms.  The contamination from the host galaxy in these regions
  is below the detection limit of our decomposition.

\item {\it Stellar content.} The stellar light of the host galaxy is
  composed mainly by giant GIII ($\sim 50$\%) and KIII ($\sim$ 35\%)
  stars with a small contribution from young O--A stars, and MIII
  giants ($<15$\%). The stellar light of the polar disk is also
  dominated by giant GIII ($\sim 45$\%) and KIII ($\sim 35$\%) stars
  with a minor contribution from young O--A stars and MIII giants
  ($<20$\%). The contribution of young stars has a radially increasing
  gradient towards outer regions, being less than $\sim$ 10\% of the
  galaxy flux in the inner 1~kpc and reaching a higher fraction in
  the outer regions (up to $\sim$ 30\% for $R> 1$~kpc).
\end{itemize}

For the first time, the spectral decomposition along the minor axis of
NGC~4650A has separated the contributions of the polar disk and the
host galaxy to the total rotation velocity and dispersion observed
along the line-of-sight. In doing so, we have addressed the
uncertainties and the ambiguities previously pointed out by
\citet{Iodice+06} (see Section \ref{sec:intro}). Specifically, we note
the following:

\begin{itemize}

\item The rotation of the host galaxy along its photometric minor axis
  is an intrinsic property; it is not an artifact caused by the
  contamination of the stars in the polar disk that lie along the same
  line of sight velocity distribution. The host galaxy minor axis
  rotation indicates that the gravitational potential is non
  axi-symmetric, suggesting triaxiality.  The triaxial scenario is
    also supported by the radial change of ellipticity and position
    angle (which are indeed indications of triaxiality,
    e.g., \citealt{Binney78, Bertola+79}) that is observed in the isophotes of
    the host galaxy \citep{Iodice+06}.

  \item Our results support the picture in which the polar disk of
    NGC~4650A formed through several accretion episodes. The
    counter-rotating nature of the polar disk indicates that the
    accreted material had different angular momenta. The ionized gas
    and the H{\small I} rotate along the same direction as the outer
    disk \citep{Whitmore+87, VanGorkom+87, Arnaboldi1997}. This
    suggests that the gas cloud that formed the stars in the outer
    disk was the latest to be accreted and that it had removed the
    pre-existing counter-rotating gas associated with the stars in the
    central 5'' \citep{Thakar+96, Pizzella+04}. This interpretation is
    consistent with the radial gradient of O-A stars in the
    disk. Indeed, the higher fraction of O-A stars in the outer
    regions of the disk with respect to the inner regions suggests
    that the accretion episodes occurred at different times with the
    stars at larger radii in the disk being formed later. The age
    difference between the central and outer regions is consistent
    with the predition of cosmological simulation, where
    counter-rotating disks form via filament accretion
    \citep{Algorry+14}.

\item The increase of velocity dispersion observed along $P.A. =
  152^{\circ}$ is an intrinsic property of the stars in the polar
  disk and not the result of the superimposition of stars from the
  two components along the line of sight. The higher velocity
  dispersion of the stars in the outer regions of the polar disk,
  combined with the larger fraction of young O-A stars, suggest that
  the outer regions of the disk formed later.

\end{itemize}

The piece of information, which is still missing, is the
three-dimensional shape of the overall dark matter halo. The
complexity of the host-galaxy kinematics revealed by our analysis
indicates that the full two-dimensional maps of the kinematics are
needed to recover it. In this framework, data collected with the
second generation integral field spectrograph MUSE at VLT
\citep{Bacon+2010} combined with our spectral decomposition analysis
will be the key (i) to further investigate higher order moments
  in the velocity distribution and their impact on the kinematics,
(ii) to recover the shape of the dark matter halo and the dark matter
content in NGC~4650A, and (iii) to confirm the triaxial shape of the
gravitational potential.

\begin{acknowledgements}
  E.I. wishes to thank the European Southern Observatory for having
  given financial support and hospitality during her visit in July
  2013, within the ESO Scientific Visitor program. The authors thank
  R. van den Bosch, O. Gerhard, N. R. Napolitano, and T. de Zeeuw for
  useful comments and discussions. The authors wish to thank the
  anonymous referee for his/her constructive feedback. This work is
  based on observations taken at the ESO La Silla Paranal Observatory
  within the Program ID~70.B-0277.
\end{acknowledgements}

\bibliography{n4650a}

\end{document}